\documentstyle[apb_cite]{mn}

\newcommand{\lta}{{\small\raisebox{-0.6ex}{$\,\stackrel
{\raisebox{-.2ex}{$\textstyle <$}}{\sim}\,$}}}
\newcommand{\gta}{{\small\raisebox{-0.6ex}{$\,\stackrel
{\raisebox{-.2ex}{$\textstyle >$}}{\sim}\,$}}}

\begin{document}

\title[Superhumps in AM~CVn]
{Superhumps, Magnetic Fields and the Mass-ratio in AM Canum Venaticorum}

\author[Pearson]
{K.\ J.\ Pearson\thanks{Send offprint requests to: 
kjp1@rouge.phys.lsu.edu}
\\
Louisiana State University, Department of Physics and Astronomy,
Nicholson Hall, Baton Rouge, LA 70803-4001, USA. \\
}

\date{Accepted . Received ; 
in original form }

\maketitle

\begin{abstract} 
We show that the observed $K$ velocities and periodicities of AM~CVn can be 
reconciled given a mass ratio $q\approx0.22$ and a secondary star with a 
modest magnetic field of surface strength $B\sim1~{\rm T}$. We see that 
the new mass ratio implies that the secondary is most likely semi-degenerate.
The effect of the field on the accretion disc structure is examined.
The theory of precessing discs and resonant orbits is generalised to encompass
higher order resonances than 3:2 and shown to retain consistency with the new 
mass ratio. 
\end{abstract}


\section{Introduction}
\label{sec:intro}

AM~CVn is the prototype of a helium-rich class of ultra-short
period Cataclysmic Variable (CV) binaries. The eleven member systems consist 
of a white dwarf primary accreting material, through
Roche lobe overflow, from a companion star that is itself degenerate
or semi-degenerate. \scite{nelemans01a} recognised two formation scenarios 
for these systems. The 
``white dwarf family'' arise from detached double degenerate white
dwarf binaries that evolved into contact through gravitational wave
radiation \cite{faulkner72}. The ``helium star family'' arise from systems 
where a low-mass helium-burning secondary is brought into contact, again 
through gravitational wave radiation. Once mass transfer has reduced 
$M_{2}<0.2M_{\odot}$, core helium burning ceases and the star becomes 
semi-degenerate \cite{savonije86,iben91}. 

Seven of the known AM~CVn stars, including the prototype, show periodicities 
in addition to their orbital modulations that are generally interpreted as 
arising from a precessing, non-axisymmetric, accretion disc. Several of the 
members show 
regular dips in brightness on a timescale of a few days. This suggests a
similar phenomenon to the disc thermal instabilities in dwarf novae 
\cite{smak83,cannizzo88,tsugawa97} albeit with a helium rather than hydrogen
dominated disc. Dwarf novae show regular outbursts where  the luminosity
increases by ~2--5 magnitudes and a subset also show ``superhumps'' caused by a
precessing accretion disc. Here, though, the default state appears to be 
``high'': ie. one in which helium is ionized and the disc has a high viscosity.
At some point the disc drops
below a critical temperature, helium recombines and the disc switches to a 
low viscosity state. Material collects in the accretion disc and switches back
to the high state at a second, critical temperature. These critical 
temperatures are normally converted to equivalent critical surface densities
$\Sigma_{\rm crit}$. Exceptions are AM~CVn itself, that has never been 
observed to dip, and GP~Com which appears to be in a permanent low state 
\cite{warner95}.

\scite{nelemans01b} examined observations of AM~CVn in detail and arrived
at a definite identification of $P_{\rm orb}=1028.73~{\rm s}$. Using 
the beat period, $P_{\rm b}=13.38~{\rm h}$, standard precessing disc theory 
gives a mass ratio for the system 
$q=0.087$. However, their measurements of the HeII 4286~\AA~line suggested 
$K_{1}=53\pm6~{\rm km}~{\rm s}^{-1}$ which, coupled with 
$K_{2}=210$--$280~{\rm km}~{\rm s}^{-1}$,
indicate a mass ratio in the range $0.19<q<0.25$. The authors noted
the inconsistency implicit here and chose to rely on the commonly used
mass-ratio inferred from disc theory. They also noted 
the alternative that the usual precession $P_{\rm b}$-$P_{\rm orb}$-$q$ 
relation (given later in equation (\ref{eqn:familiar})) 
might be inappropriate. 

The purpose of this 
paper is to reconcile these two methods of determining $q$  by proposing that 
a modest secondary magnetic field is present in AM~CVn. We suggest that the 
true AM~CVn mass ratio is that derived directly 
from the $K$ velocities and that the widely accepted
$P_{\rm b}$-$P_{\rm orb}$-$q$ relation used by \scite{nelemans01b} 
is indeed inappropriate in this case. 

\scite{pearson97} studied the effect of a secondary's field on the behaviour
of dwarf novae: placing upper limits of $B_{\rm L1}<0.16~{\rm T}$ 
($\mu_{2}\lta7\times10^{34}~{\rm G~cm}^{3}$) in U~Gem and 
$B_{\rm L1}<0.023~{\rm T}$ ($\mu_{2}\lta4\times10^{32}~{\rm G~cm}^{3}$) in 
Z~Cha. They found that stronger fields had the effect of shrinking the disc
radius and of exciting resonances in the disc. However, the disc was too 
small to access the 3:2 resonance normally observed in superhumping systems
and instead higher order resonances appeared. These higher resonances are 
available, in principle, to all CV accretion discs that do not reach the 
required 3:2 resonant radius, even in the absence of a secondary field. 
However, in practice, the 
tidal effect that drives the superhump resonance is unable to excite 
the higher orders before they are damped out by viscosity. It is the addition 
of the secondary field that enables these higher resonances to be excited
and their appearance or otherwise that enables us to constrain
the secondary's magnetic field strength. We would expect 
a secondary with a magnetic field in AM~CVn to have a similar effect: ie. to 
shrink the disc, excite higher order resonances and to thereby alter the 
applicable $P_{\rm b}$-$P_{\rm orb}$-$q$ relation.

It has long been problematic to understand the origin of the 
magnetic field in the secondaries of CVs: particularly those systems that lie 
below the $2$--$3$~hr ``period
gap'' where the stars are fully convective. However,
we know that such a field must be possible, if not ubiquitous, from 
the fact that the Polars are able to lock in synchronous rotation through
the interaction of the two stellar dipoles. This locking requires that 
the secondary has a magnetic moment 
$\mu_{2}\sim10^{33}$--$10^{34}~{\rm G}~{\rm cm}^{3}$ 
\cite{king90} that is similar to that of the primary. Observationally, 
results from a variety of open clusters 
\cite{jones96,stauffer97,terndrup00,reid00}
suggest that the rate of magnetic braking does not reduce abruptly
when stars become fully convective ($M\lta 0.3M_{\odot}$); implying
the continued presence of a significant magnetic field.
Such a field also has implications for the standard model of the 
period gap that relies on the rate of magnetic braking 
dropping sharply at the upper end \cite{andronov03}. 

Recently, models have begun to emerge for the generation of magnetic 
fields by fully convective, rapidly rotating, young stars \cite{kuker99} that 
may survive as 
relics into the main sequence lifetime \cite{kitchatinov01}. These fields are  
non-dipolar and highly non-axisymmetry. 
Such geometries, if repeated in fully convective CV secondaries, may have 
interesting consequences for the equilibrium alignment of Polars and the 
interpretation of the ``hot spot'' to which
material is channelled at the white dwarf surface. 
These models, however, have not so far 
extended into the extremely rapid rotation regime required for CV secondaries;
especially with as short a $P_{\rm orb}$ as AM~CVn.
We are also confronted, in this particular case, with a secondary
that is at least semi-degenerate with the unknown 
consequences for the field that such a structure and previous evolution will 
have. However, since for a dipolar field $f\propto B^2\propto r^{-6}$, 
it would 
require a relatively much weaker dipole moment to achieve the same
local field strength in these short period, small separation systems than in 
wider long period CVs. It is, therefore, not unreasonable to expect to see 
effects in these systems that might not be apparent in their longer period 
counterparts. 

In section~\ref{sec:theback} we outline the important background theory  to the
problem; generalising the theory of superhump resonances to higher orders
in section~\ref{sec:supthry} and deriving component masses for the system
in section~\ref{sec:masses}. We describe simulations of the system in 
section~\ref{sec:sims}.

\section{Theoretical Background}
\label{sec:theback}
\subsection{Predicting Superhump Periods}
\label{sec:supthry}

\scite{hirose90} derived the general expression for the ratio of the 
disc precession $\omega_{\rm p}$ and orbital $\omega_{\rm orb}$ frequencies 
in terms of the mass ratio and radius of disc material.
Their equation (8) is
\begin{equation}
\frac{\omega_{\rm p}}{\omega_{\rm orb}} =
\frac{q}{\left(1+q\right)^{\frac{1}{2}}}
\left[\frac{1}{2r^{\frac{1}{2}}} \frac{\rm d}{{\rm d}\!r}
\left(r^{2} \frac{{\rm d}\!B_{0}}{{\rm d}\!r}\right)\right],
\label{eqn:osaki8}
\end{equation}
where
\begin{equation}
B_{0}(r) = \frac{1}{2}b^{0}_{\frac{1}{2}} = F\left(\frac{1}{2},\frac{1}{2},1,
r^{2}\right)
\end{equation}
\cite{brumberg95} is the zeroth order Laplace coefficient given in terms of 
the hypergeometric function $F$, $q=M_{2}/M_{1}~(<1)$ is the mass-ratio
and $r$ is the radius of 
orbiting material expressed as a fraction of the separation $a$. 
Writing the hypergeometric function in its series form
\begin{eqnarray}
F\left(\frac{1}{2},\frac{1}{2},1,r^{2}\right) & = & \sum_{n=0}^{\infty}
\frac{\left(\frac{1}{2}\right)_{n}\left(\frac{1}{2}\right)_{n}}
{\left(1\right)_{n} n!}r^{2n} \\
&=& 1 + \frac{1}{4} r^{2} + \frac{9}{64} r^{4} + ...+C_{n}r^{2n}.
\label{eqn:fseries}
\end{eqnarray}
The coefficients are calculable exactly but rapidly become unwieldy with large 
numerators and denominators. However, after some algebra we can rewrite
the general expression for them as
\begin{eqnarray}
C_{n} & = & \left[\frac{ (1)(3)(5)...(2n-1)}{n! 2^{n}}\right]^{2}\\
& = &\prod_{m=1}^{n} \left(\frac{2m-1}{2m}\right)^{2},
\label{eqn:fcoeffs}
\end{eqnarray} 
which at least clears away the less transparent fractional terms in
Pochhammer's symbol of $\left(\frac{1}{2}\right)_{n}$.  
Combining equations (\ref{eqn:osaki8}), (\ref{eqn:fseries}) and 
(\ref{eqn:fcoeffs}), and after some more algebra and a little differentiation, 
we can arrive at a general relation for resonant orbits 
\begin{eqnarray}
\frac{\omega_{\rm p}}{\omega_{\rm orb}} & = &
\frac{3}{4}\frac{q}{\left(1+q\right)^{\frac{1}{2}}}
r^{\frac{3}{2}}\left[1+c_{2}r^{2}+c_{3}r^{4}+c_{4}r^{6}+...\right. \nonumber \\
& & +
\left. c_{n}r^{2(n-1)}
\right],
\end{eqnarray}
where the coefficients are related by
\begin{equation}
c_{n}=\frac{2}{3}(2n)(2n+1)C_{n}.
\label{eqn:coeffrel}
\end{equation}
The values of $C_{n}$ and $c_{n}$ up to $n=5$ are listed in 
Table~\ref{tab:coeffs} for convenience.

\begin{table}
\begin{center}
\begin{tabular}{lcccccc}
\hline
$n$     & $0$ & $1$ &  $2$ &  $3$ &  $4$ &  $5$ 
\\ \hline
$C_{n}$ & $1$ & $\frac{1}{4}$ & $\frac{9}{64}$ & $\frac{25}{256}$ & 
$\frac{1225}{16384}$ & $\frac{3969}{65536}$ 
\\
\\
$c_{n}$ & $0$ &  $1$ & $\frac{15}{8}$ & $\frac{175}{64}$ &
$\frac{3675}{1024}$ &$\frac{72765}{16384}$ 

\\
\hline

\end{tabular}
\end{center}
\caption{Coefficients contained in the expansions defined in equations 
(\protect\ref{eqn:fcoeffs}) and (\protect\ref{eqn:coeffrel}) up to $n=5$.}
\label{tab:coeffs}
\end{table}

\begin{figure}
\vspace{6.25cm}
\includegraphics{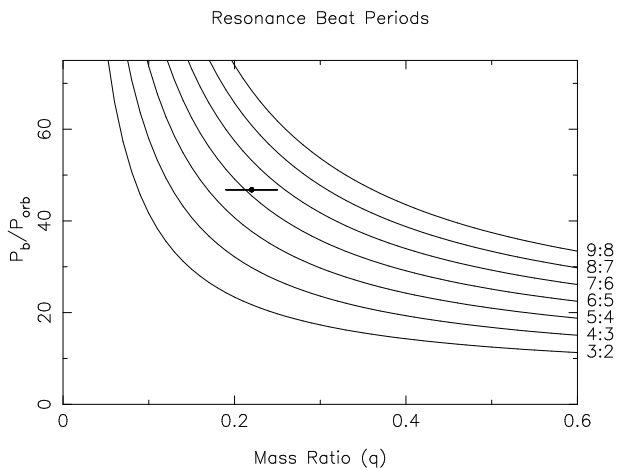}
\caption{Relation between the precession to orbital beat period and
mass ratio given in equation~(\protect\ref{eqn:final}) for different 
resonances $j$:$j-1$. Also indicated is the range of the mass
ratio allowed by $K$~velocity measurements for AM~CVn.}
\label{fig:orbq}
\end{figure}

The radii
of resonant orbits is given by \scite{FKR}. Restricting ourselves to the
case of $j$:$j-1$ resonances, that are strongest, we obtain 
\begin{equation}
r_{j}
=\frac{1}{j^{\frac{2}{3}}\left(1+q\right)^{\frac{1}{3}}}.
\end{equation}
Thus we finally arrive at
\begin{eqnarray}
\frac{\omega_{\rm p}}{\omega_{\rm orb}} & = &
\frac{3}{4j}\frac{q}{1+q}\left[1+
\frac{c_{2}}{j^{\frac{4}{3}}(1+q)^{\frac{2}{3}}}+
\frac{c_{3}}{j^{\frac{8}{3}}(1+q)^{\frac{4}{3}}}+
\right. \nonumber\\ & & \left. 
\frac{c_{4}}{j^{4}(1+q)^{2}}+
\frac{c_{5}}{j^{\frac{16}{3}}(1+q)^{\frac{8}{3}}}+...\right].
\label{eqn:prefinal}
\end{eqnarray}
Since these are resonant orbits, the (long) beat period is related to the 
orbital and precession angular frequencies by
\begin{equation}
P_{b}=P_{\rm orb}\frac{j}{j-1}\frac{\omega_{\rm orb}}{\omega_{\rm p}}.
\label{eqn:final}  
\end{equation}
This is plotted for several resonances in Figure~\ref{fig:orbq}.
The familiar approximation
\begin{equation}
P_{\rm b}=A \frac{1+q}{q} P_{\rm orb},
\label{eqn:familiar}
\end{equation}
where $A \approx 3.85$ for $0.1<q<0.22$
\cite{warner95}, is recovered by setting $j=3$ and evaluating the term in 
square brackets in (\ref{eqn:prefinal}) with $q=0.16$. The limiting mass 
ratio $q\approx0.22$ found by \scite{whitehurst88a} arises from the largest 
value for which $r_{3}$ remains within the last stable stream line
\cite{molnar92}.

We can derive parallel approximations to (\ref{eqn:familiar}) for other 
resonances in a similar way. The coefficient $A$ need only be recalculated 
with the appropriate value of $j$. Table~\ref{tab:rescon} gives $A$ for 
different values of $j$; appropriate again to the range $0.1<q<0.22$ by 
evaluating with $q=0.16$.

\begin{table}
\begin{center}
\begin{tabular}{lccccccccc}
\hline 
$j$ & $3$ & $4$ & $5$ & $6$ & $7$ & $8$ & $9$ \\
$A$ & $3.86$ & $5.33$ & $6.75$ & $8.16$ & $9.55$ & $10.9$ & $12.3$ \\
\hline
\end{tabular}
\end{center}
\caption{Coefficient $A$ contained in the approximation 
(\protect\ref{eqn:familiar}) for several resonances and appropriate to
 $0.1<q<0.22$.}
\label{tab:rescon}
\end{table}

For the particular case of AM~CVn, using $q=0.22$ the full expressions 
give rise to the predicted superhump periods and disc radii shown in 
Table~\ref{tab:amsh}. 

\begin{table}
\begin{center}
\begin{tabular}{ccc}
\hline
Resonance  & Radius & $\frac{P_{\rm b}}{P_{\rm orb}}$ \\
($j:j-1$)  & ($r_{j}$)  & \\
\hline
$3:2$  &  $0.450$     &  $21.8$ \\
$4:3$  &  $0.371$     &  $29.8$ \\
$5:4$  &  $0.320$     &  $37.7$ \\
$6:5$  &  $0.283$     &  $45.5$ \\
$7:6$  &  $0.256$     &  $53.2$ \\
$8:7$  &  $0.234$     &  $60.8$ \\ 
$9:8$  &  $0.216$     &  $68.4$ \\\hline
\end{tabular}
\end{center}
\caption{The radius and beat period of superhump resonances for $q=0.22$.}
\label{tab:amsh}
\end{table}

\subsection{Mass Determinations}
\label{sec:masses}

We can write the approximation by \scite{eggleton83} for the size of the
secondary's  Roche lobe as
\begin{equation}
R_{2} = 
\left[\frac{G}{4\pi^{2}}\right]^{\frac{1}{3}} 
\frac{0.49 q^{\frac{1}{3}} (1+q)^{\frac{1}{3}}}
                    {0.6q^{\frac{2}{3}} + \ln(1+q^{\frac{1}{3}})} 
M_{2}^{\frac{1}{3}} P_{\rm orb} ^{\frac{2}{3}},
\end{equation}
where we have eliminated $a$ using Kepler's Third Law. Knowing $P_{\rm orb}$
and $q$, for a given mass-radius relation we can eliminate $R_{2}$ and 
solve for $M_{2}$.

For a fully degenerate He white dwarf secondary, the relation
\begin{equation}
\frac{R}{R_{\odot}}\approx
0.0106-0.0064 \ln\left(\frac{M}{M_{\odot}}\right) + 0.0015 \left(
\frac{M}{M_{\odot}} \right)^{2}
\label{eqn:eosdegen}
\end{equation}
\cite{zapolsky69,rappaport84} is appropriate. For a semi-degenerate secondary, 
the relation is more problematic. However, approximations of the form
\begin{equation}
\frac{R}{R_{\odot}}\approx b\left(\frac{M}{M_{\odot}}\right)^{-\alpha}
\label{eqn:eossemi}
\end{equation}
have been found by \scite{tutukov89} with $b=0.043, \alpha=0.062$ (hereafter 
`TF' parameters) and \scite{savonije86} with $b=0.029, \alpha=0.19$ (hereafter
`SKH'). The resulting values for $M_{1}$ and $M_{2}$  are listed in
Table~\ref{tab:masses} for each case.

\begin{table}
\begin{center}
\begin{tabular}{rccc}
\hline
Secondary Type & $M_{1}/M_{\odot}$ & $M_{2}/M_{\odot}$  \\
\hline
Degenerate & 0.156 & 0.034 & \\
Semi-degenerate (TF) & 0.529 & 0.116 & \\
Semi-degenerate (SKH) & 0.418 & 0.092 & \\ \hline
\end{tabular}
\end{center}
\caption{Derived values for the AM~CVn component masses assuming $q=0.22$
and different mass-radius relations for the secondary.}
\label{tab:masses}
\end{table}

\scite{nelemans01a} carried out a population synthesis study of AM~CVn
systems. They noted that the previously adopted $q=0.087$ gave rise to
component masses that were difficult to reconcile with their results. 
For a degenerate secondary and that mass ratio, the system would lie at the 
low end of the 
predicted distribution and for a semi-degenerate system would imply a primary
mass close to, if not in excess of, the Chandrasekhar mass. Locating the 
results from Table~\ref{tab:masses} on their Figure~1, we see that for 
$q=0.22$ and a fully degenerate secondary,
the system would have to have been born at an implausibly unlikely position
in the tail of the distribution. For a semi-degenerate secondary, however, the 
difficulties of a very high $M_{1}$ have now been resolved with a
more comfortable $M_{1}\sim0.5M_{\odot}$.
  
\section{Simulations}
\label{sec:sims}

Modelling was carried out using the same code HYDISC 
\cite{whitehurst87,whitehurst88b} used originally to demonstrate the tidal
origin of superhump resonances \cite{whitehurst88a}. It has 
since been adapted to a variety of other CV \cite{wynn95,wynn97,king99} and 
young star \cite{pearson95,ultchin02} applications.
The modifications necessary to include a secondary magnetic field were
discussed in \scite{pearson97}. 

Simulations of the effect of the secondary field in CVs were recently 
reexamined by \scite{murray02} using SPH. These authors modelled the magnetic 
interaction through an acceleration
\begin{equation}
{\bf a} = -k[{\bf v} - {\bf v}_{f}]_{\perp},
\label{eqn:dragform}
\end{equation}
where ${\bf v}$ and ${\bf v}_{f}$ are the velocities of the material and field 
lines respectively and  $\perp$ indicates that only components 
perpendicular to the field lines are considered.
Such a form, originating from models that used a diamagnetic 
prescription for the interaction, can also be applied to more general forms. 
In that case, our ignorance of the details of
the interaction are contained in $k$ and, in particular, in its 
behaviour as a function of position and magnetic field strength. We prefer
here to retain the model of the interaction contained in
\scite{pearson97}. That is, that the disc is assumed to wind up the magnetic
field through advection until $B_{\phi}\sim B_{z}$ \cite{aly90,wang96}. This 
enables us to maintain
an explicit relationship between the force and secondary field strength.
Specifically, this earlier study showed that the acceleration 
experienced by each particle could be modelled as
\begin{equation}
{\bf a} = 
- \frac{2 \pi B^{2} R \Delta\!R}{\mu_{0} \beta N m_{\rm p}} {\bf e}_{\phi}
\label{eqn:amag}
\end{equation}
where $m_{\rm p}$ is the mass represented by each particle, $N$ is the number 
of particles in a bin of width $\Delta\!R$ at a distance $R$ from the white 
dwarf, $B$ is the local magnetic field strength and $\beta$ is a dimensionless
parameter related to the exact geometry of the field distortion (set equal
to unity hereafter). Since we have a strong constraint in the superhump
period, we can place limits on the possible secondary field strength.

The secondary's magnetic field acts on the disc material in a similar way to 
the gravitational tidal interaction. However, the disc material is now coupled 
more strongly to the secondary by the field. As a result, the disc needs to 
find a new configuration where the outward advection of angular momentum can
meet the more efficient transfer back to the secondary. The 
$B^{2}\propto r^{-6}$ dependence provides a sharp ``spike'' to the 
interaction that acts to promote the appearance of resonance phenomena. 
There is a limiting field strength below which the disc can find a suitable 
new structure. Above this, angular momentum extraction is too efficient,the 
disc is completely disrupted and material is rapidly accreted by the primary. 

We adopted the system orbital period of 1028.7~s confirmed by 
\scite{nelemans01b}, the mass ratio $q=0.22$ in the middle of the range implied
by the $K$ velocities and $M_{1}=0.5 M_{\odot}$.

Each simulation was started with a burst of particles over an initial 
$2P_{\rm orb}$ before being reduced to a steady $90$ per orbit. The magnetic
interaction was ``switched on'' after $3P_{\rm orb}$ once a disc had begun to
form. Each of the simulations was allowed to run for at least $500P_{\rm orb}$,
allowing an equilibrium to be reached.

We can estimate the effective value of the Shakura-Sunyaev parameter $\alpha$
from the ring of material that forms initially at the
circularization radius $R_{\rm c}$ in a non-magnetic simulation. Ignoring 
any additional mass transfer of material, the surface density $\Sigma$ of an 
initial ring of mass $m$ will evolve as 
\begin{equation}
\Sigma(x,\tau)=\frac{m}{\pi R_{\rm c}^{2}} \tau^{-1} x^{-\frac{1}{4}}
\exp\left\{-\frac{(1+x^{2})}{\tau}\right\} I_{\frac{1}{4}}
\left(\frac{2x}{\tau}\right)
\end{equation} 
\cite{FKR} where $x=R/R_{\rm c}$, $I$ is a modified Bessel function and
$\tau=12 \nu t R_{\rm c}^{-2}$ . Fitting this to the early time behaviour 
where the much 
lower ``steady state'' mass transfer has negligible effect gives a viscosity
$\nu\approx6.1\times10^{8}~{\rm m}^{2}~{\rm s}^{-1}$. From a Gaussian fitted
to the vertical disc profile we have a scale height 
$H\approx1.2\times10^{5}~{\rm m}$. Hence, using a sound speed appropriate to a
fully ionized gas of solar abundance at $\sim10^{4}~{\rm K}$, we derive 
$\alpha\sim0.3$.

The particle mass used in the magnetic interaction 
(\ref{eqn:amag}) was determined from a nominal
mass transfer rate of $10^{14}~{\rm kg}~{\rm s}^{-1}$. However, as
the particles have negligible mass gravitationally, this is the only place in 
which $m_{\rm p}$ appears in the equations of motion. Thus the magnetic 
simulations may be characterised
by the parameter $\frac{B^{2}}{\dot{M_{2}}}$. We use
the L$_{1}$ point as the fiducial position for the surface strength of the 
secondary's field and assume the field strength to vary as for a dipole. 

\begin{figure}
\vspace{6.25cm}
\includegraphics{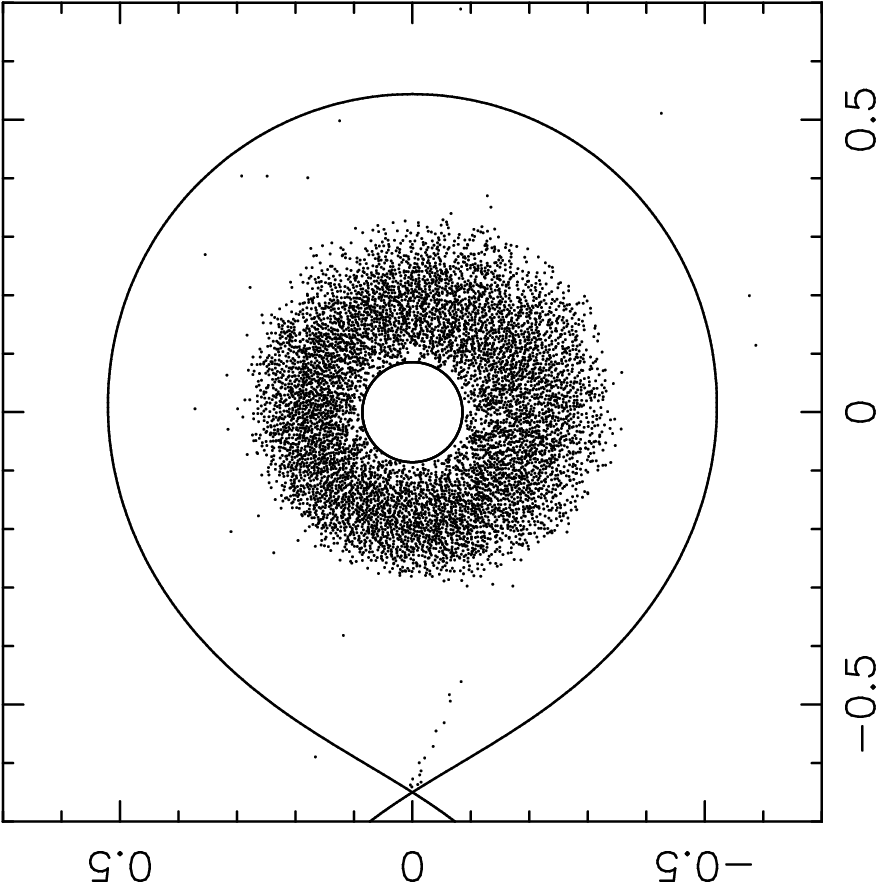}
\caption{Particle distribution for the $B_{\rm L1}=0.7~{\rm T}$
simulation showing the precessing, elliptical disc pattern. 
}
\label{fig:sim}
\end{figure}

\begin{figure}
\vspace{6.25cm}
\includegraphics{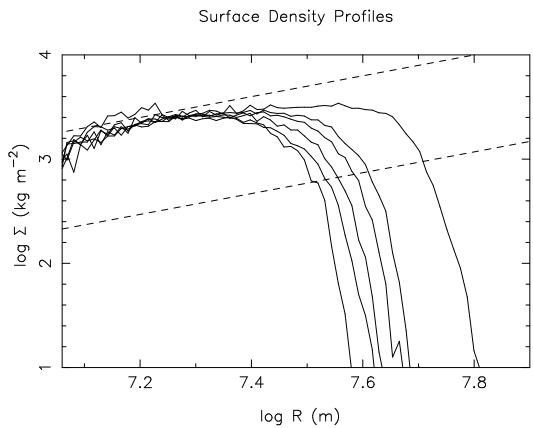}
\caption{Mean surface density profile for equilibrium simulations with 
$B_{\rm L1}=0,0.6,0.65,0.7,0.75,0.8$~T (decreasing radius). 
Also shown (dashed line) are the critical surface densities for transition
to the ``hot'' (upper line) and ``cold'' states.}
\label{fig:surdens}
\end{figure}

In common with \scite{pearson97} and \scite{murray02}, we see that the 
magnetic field tends to shrink and promote resonant behaviour in the disc
(see eg. Figure~\ref{fig:sim})
although the resonances are relatively weak compared to SU~UMa superhumps.
Simulations with $B_{\rm L1}\lta0.5~{\rm T}$ are unable to excite any 
resonances.
Similarly, when $B_{\rm L1}\gta1~{\rm T}$ the disc is completely disrupted.
Surface density profiles are shown in Figure~\ref{fig:surdens} for
different magnetic field strengths. The critical surface densities
for ``dwarf nova'' type thermal instabilities in helium discs are taken
from \scite{tsugawa97} using a hot state $\alpha=0.3$ and cold state
$\alpha=0.03$. We see that all the ``magnetic'' discs have 
surface densities between the two critical values and, hence, that they 
can maintain either a ``hot'' or ``cold'' state. 
However, the effective value of $\alpha$ derived above implies that the
simulations represent a ``hot'' disc configuration which is consistent with 
the observation that AM~CVn is in a permament ``high'' state.

\begin{figure}
\vspace{6.25cm}
\includegraphics{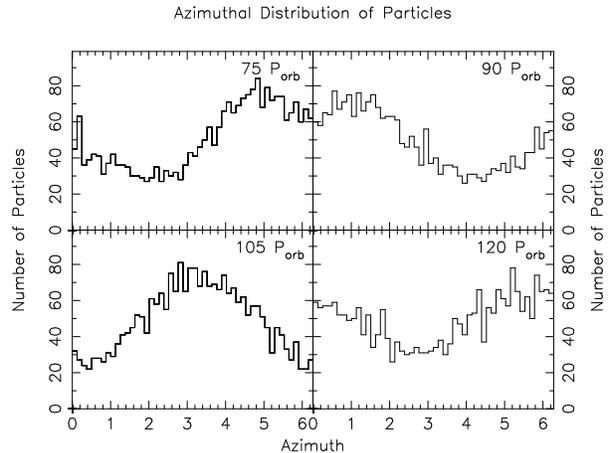}
\caption{Angular distribution of particles in the disc for the 
$B_{\rm L1}=0.7~{\rm T}$ simulation with $0.225<r<0.325$
at $t=75,90,105,120~P_{\rm orb}$. The profile can be seen to drift slowly 
towards higher azimuthal angles.}
\label{fig:aprof}
\end{figure}

We can derive the precession period of the disc
by considering plots of the angular distribution of particles, such as 
those in Figure~\ref{fig:aprof}, plotted every 
$P_{\rm orb}$. We can also use Fourier transforms 
of the number of particles in a particular azimuthal bin, although, even 
running for $512~P_{\rm orb}$ there is limited resolution at long $P_{\rm b}$. 
These periods are summarised for each field strength in 
Table~\ref{tab:results}. The models bracket the observed 
$P_{\rm b}=46.8P_{\rm orb}$ and, conservatively, place limits on the 
allowed field strength of
$0.65~{\rm T}< B_{\rm L1}\left(\frac{\dot{M_2}}{10^{14}~{\rm kg}~{\rm s}^{-1}}
\right)^{-1/2}< 0.8~{\rm T}$.
These are equivalent to a magnetic moment, with the assumed mass transfer
rate, of
$(6.0$--$7.3)\times10^{32}~{\rm G~cm}^{3}$ which is at the low end of the 
range deduced for Polar secondaries.
We do not quote formal errors on these limits since they are dominated 
by the systematic effects in our assumptions regarding the geometry
of the field distortion. \scite{pearson97} estimated 
that in a worst case these values could be in error by no more than a factor 3
and were probably significantly better.

\begin{table}
\begin{center}
\begin{tabular}{cccccc}
\hline
$B_{L1}~(T)$  & 0.6  &   0.65 &   0.7  &    0.75 &  0.8 \\ \hline
$\frac{P_{\rm b}}{P_{\rm orb}}$ & 28 & 35 & 43 & 51 & 58 \\ 
$j$ & 4 & 5 & 6 & 7 & 8\\ 
\hline
\end{tabular}
\end{center}
\caption{Measured beat period and closest predicted resonance for different
field strengths.}
\label{tab:results}
\end{table}

\section{Summary}

We have shown that the observed $K$ velocities in AM CVn can be reconciled  
with our understanding of superhump resonance behaviour given a modest 
secondary magnetic field. The field acts to shrink the disc and excite higher 
order resonances than those tidally excited in SU~UMa systems. The sensitivity
of the disc behaviour to the magnetic field places a tight 
constraint upon the possible field strength of the secondary although our 
errors are dominated by assumptions inherent in the model. A value of
$q=0.22$ strongly suggests that the secondary in AM~CVn is semi-degenerate
and is more easily reconciled with population synthesis results than the
previous $q=0.087$.

We intend to return to this system in a future paper exploring more 
exhaustively the possible
parameter space ($M_{1}$, $q$, $B$, $\nu$ etc.) which AM~CVn can occupy and 
still
reconcile the disc precession and $K$ velocity measurements. Further direct
measurements to confirm $K_{1}$ would be extremely useful.  

\section*{ACKNOWLEDGEMENTS}

The author would like to thank the referee Graham Wynn for his helpful remarks
on the first version of the paper and also both Juhan Frank and Keith Horne for
insightful comments on a draft version. This work has been supported, in 
part, by the U.S. National Science Foundation through grant AST-9987344 and, 
in part, through LSU's Center for Applied Information Technology and Learning.
\\

\end{document}